\documentclass[12pt,a4paper]{article}

\usepackage{graphicx,graphics,color,epsfig}
\usepackage{latexsym}
\usepackage{amsmath,amsfonts,amssymb}
\usepackage{verbatim}

\begin{document}

\begin{center}
\Large {\bf Light sterile neutrinos effects in processes with electron and
muon neutrinos}
\vskip 8mm
\large
V. V. Khruschov, S. V. Fomichev
\vskip 5mm
\large
{{\it NRC Kurchatov Institute, 123182 Moscow, Russia }} \\
\end{center}

\vskip 6mm

\begin{abstract}
\noindent
Sterile neutrinos with various masses could be participated in astrophysical
and cosmological processes including mixing with active neutrinos, if exist.
It is considered the possible effects of mixing of active and sterile
neutrinos with masses of the order or less than 1 eV. For oscillation
processes involving electron and muon neutrinos, as well as for beta decay and
neutrinoless double beta decay processes, the contributions of light sterile
neutrinos in the characteristics of these processes are calculated. For this
purpose the estimates of the mixing parameters in the model with three active
and three sterile neutrinos are made taking into account experimental data.
Two cases of sterile neutrinos masses distribution are considered in detail.
The results obtained can be used for interpretation of available experimental
data, and also for predictions of subsequent experimental results.
\end{abstract}
\vspace*{6pt}


\bigskip


\vskip 8mm

\section{Introduction}
\label{Section_Introduction}

It is well known that in the framework of the Standard Model (SM) of
electromagnetic, weak and strong interactions of particles a quantitative
agreement between most of experimental and theoretical results was achieved
\cite{pdg}. However, some data are emerging and increasing in number that
cannot be well described within the framework of the SM. For instance, it
relates to oscillations of active neutrinos, which have been detected
experimentally. Nonzero neutrino masses must be invoked for their explanation.
In this case SM is called as the modified SM ($\nu$SM). Using the neutrino
mass states makes it possible to explain the oscillations of the known
neutrino flavor states, i.e. electron, muon and tau neutrinos ($\nu_e$,
$\nu_{\mu}$, $\nu_{\tau}$) with the help of the
Pontecorvo--Maki--Nakagawa--Sakata mixing matrix $U_{\rm PMNS}\equiv U$. The
standard parametrization for matrix $U$ is given in the review \cite{pdg}.

In the present paper some peculiarities of neutrino processes are considered.
For their explanation possibly it is necessary to go out beyond not only the
SM but also the $\nu$SM. This is especially true in regard to so called
neutrino anomalies at short distances (short baselines, SBL) from the source
\cite{abaz,boser,acero}. The appearance of these anomalies can be explained
perhaps with the effects of new particles, namely, light sterile neutrinos
(LSN) with the characteristic mass scale about 1~eV. Let us mention that three
sterile neutrinos with masses of the order of 1~eV were already introduced in
Ref. \cite{AdeGo} for the explanation of the LSND anomaly. Generally a LSN
number can be arbitrary. The most using model now is the (3+1) model with one
LSN, but the $(3+2)$ and $(3+3)$ models are used as well (see, for example,
\cite{Kopp2013,Conrad2013}). Below we consider the LSN effects in processes
with electron and muon neutrinos in the framework of the version of the
$(3+3)$ model, which was elaborated in Refs.~\cite{khfo,KhFoSe,KhFom}.

The content of the paper is as follows. Section~\ref{Section_anomal} provides
short reference to SBL neutrino anomalies perceived in a number of
experiments \cite{acero}. Section~\ref{Section_OscillationModel} contains
brief description and some results of the used $(3+3)$ model with three LSN
\cite{KhFom}. In Section~\ref{ModelParameters} the test values of the model
parameters are proposed taking into account experimental data. Thereon
calculations of the survival probabilities of electron and muon neutrinos and
probabilities of transition of muon neutrinos to electron neutrinos with the
help of the obtained parameter values are carried out. The calculation results
are represented in the graphic form (Figs.~\ref{fig1}, \ref{fig2} and
\ref{fig3}). The effective masses of electron neutrinos, which can be measured
in experiments on beta decay and neutrinoless double beta decay, are also
estimated. In the final Section~\ref{Section_Conclusion} the main results of
the paper are discussed.

\section{Neutrino data anomalies on small distances from neutrino sources}
\label{Section_anomal}

In addition to the known standard data on neutrino oscillations for three
active neutrinos, indications are obtained related to anomalous data for SBL
neutrino fluxes in a number of processes. These anomalies cannot be explained
by using the oscillatory parameters for only known active neutrinos. They
include the LSND (or accelerator) anomaly (AA)
\cite{Atha1996,Agu2013,Agu2018,Denton}, the gallium (or calibration) anomaly
(GA) \cite{Abdu2009,Kae2010,Giunti2013} and the reactor (antineutrino) anomaly
(RA or RAA) \cite{Mu2011,Hu2011,Me2011,Ko,ale18,ser}. AA, GA and RA manifest
themselves at small distances, more precisely, at such distances from the
source $L$, when the value of the parameter $\Delta m^2 L/E$ is of the order
of unity (where $E$ is the neutrino energy and $\Delta m^2$ is the square of
the characteristic mass scale of the considered oscillations). $\Delta m^2$ is
equal to the difference of the squared masses of the participating neutrinos.
Excluding purely experimental problems, the SBL anomalies can be explained by
the presence of at least one neutrino with a mass of about 1 eV, which does
not interact directly with the $\nu$SM gauge bosons, therefore they are called
LSN.

AA was noticed firstly by the LSND collaboration in the reaction with the
transition of a muon antineutrino into an electron antineutrino
\cite{Atha1996}. Then this result was confirmed and extended by adding the
results of the reaction with the transition of a muon neutrino into an
electron neutrino in the MiniBooNE experiment \cite{Agu2013,Agu2018} and with
less significance in the MicroBooNE experiment \cite{Denton,agu22}. GA was
discovered during the calibration of detectors for the Ga-Ge experiment at
the SAGE and GALLEX facilities \cite{Abdu2009,Giunti2013} and has been
confirmed in the BEST experiment \cite{bar22,bar22a}. Now the confidence level
of the AA and GA in the mentioned experiments lies in the $4-5\sigma$ CL
interval \cite{agu22,bar22}.

After recalculating the values of the antineutrino flux from the reactor, the
theoretical values turned out to be $3\%$ higher than those used before
\cite{Mu2011,Hu2011} that led to RA about the $3\sigma$ level \cite{Me2011}.
However, it should be noted that the $\beta$-spectra of the decay products of
uranium and plutonium isotopes introduce large systematic uncertainties into
the reactor spectra \cite{kop21}.

\section{Some propositions and results of the (3+3) model}
\label{Section_OscillationModel}

In spite of that the SBL anomalies can be explained, as mentioned above, with
the presence only one LSN with the characteristic mass scale about 1~eV, the
number of additional sterile neutrinos with different masses, in principle, 
can be arbitrary \cite{abaz,Bilenky1977,Bilenky}. Phenomenological models with 
$N$ SN are usually denoted as ($3+N$) models.

($3+N$) models are often used to describe SBL anomalies as well as some 
astrophysical data \cite{abaza2}. It is desirable that 
the $N$ number would be minimal that is why $(3+1)$ and $(3+2)$ models
are mostly used \cite{Kopp2013}. However, taking into account the possible
left-right symmetry of weak interactions, $(3+3)$ models attract considerable
attention (see, e.g. \cite{Conrad2013,Zysina2014}). In this paper, to take
into account the LSN effects, the $(3+3)$ model is also used \cite{KhFom},
which includes three known active neutrinos $\nu_a$ ($a=e,\mu,\tau$) and three
new (in this case light sterile) neutrinos: sterile neutrino $\nu_s$, hidden
neutrino $\nu_h$ and dark neutrino $\nu_d$. Thus, the model contains six
neutrino flavor states and six neutrino mass states, therefore a
$6\!\times\!6$ mixing matrix is used. This matrix is dubbed as the generalized
mixing matrix or the generalized Pontecorvo--Maki--Nakagawa--Sakata matrix
$U_{\rm GPMNS}\equiv U_{\rm mix}$ \cite{khfo}.

$U_{\rm mix}$ can be represented as the matrix product $V\!P$, where $P$ is a
diagonal matrix containing the Majorana CP-phases $\phi_i$, $i=1,\dots,5$,
that is $P={\rm diag}\{e^{i\phi_1},\dots,e^{i\phi_5},1\}$. Below we will use
only some particular forms of matrix $U_{\rm mix}$.
In this case, we will denote the Dirac CP-phases as $\delta_i$ and $\kappa_j$,
and the mixing angles as $\theta_i$ and $\eta_j$. In doing so,
$\delta_1\equiv\delta_{\rm CP}$, $\theta_1\equiv\theta_{12}$,
$\theta_2\equiv\theta_{23}$ and $\theta_3\equiv\theta_{13}$. Only the normal
order (NO) of the active neutrino mass states and the value
$\delta_{\rm CP}=1.2\pi$ will be considered.

For compactness of formulas, we introduce symbols $\nu_b$ and $\nu_{i'}$ for
sterile left flavor fields and sterile left mass fields, respectively. So
fields $\nu_b$ with index $b$ contain fields $\nu_s$, $\nu_h$ and $\nu_d$,
while $i'$ denotes a set of indices $4$, $5$ and $6$. A total $6\!\times\!6$
mixing matrix $U_{\rm mix}$ can be represented in the form of $3\!\times\!3$
matrices $R$, $T$, $V$ and $W$:
\begin{equation}
\left(\begin{array}{c}\nu_a\\ \nu_b \end{array}\right)=
U_{\rm mix}\left(\begin{array}{c}\nu_i\\ \nu_{i'}\end{array}\right)\equiv
\left(\begin{array}{cc}R&T\\ V&W\end{array}\right)
\left(\begin{array}{c}\nu_i\\ \nu_{i'}\end{array}\right).
\label{eq_Umix}
\end{equation}
Let us represent the matrix $R$ in the form of $R=\varkappa U_{\rm PMNS}$,
where $\varkappa=1-\epsilon$, and $\epsilon$ is a small quantity. The matrix
$T$ in the equation~(\ref{eq_Umix}) must also be a small matrix as compared
with the Pontecorvo--Maki--Nakagawa--Sakata $3\!\times\!3$ matrix for active
neutrinos $U_{\rm PMNS}\equiv U$ ($UU^+=I$). So, active neutrinos mix by means 
of the $U$ matrix,
as it should be in the $\nu$SM, when choosing the appropriate normalization.
In the present state of the art, it is enough to restrict ourselves only to a
minimal number of parameters of matrix $U_{\rm mix}$, that allows one to
interpret available (still rather heterogeneous) experimental data. The
transition to the full matrix with all parameters should be done later on,
when additional data related to the SBL anomalies will be obtained.

We choose $T$ in the form of $T=\sqrt{1-\varkappa^2}\,a$, where $a$ is an
arbitrary unitary $3\!\times\!3$ matrix ($aa^+=I$), then $U_{\rm mix}$ can be
written in the following form:
\begin{equation}
U_{\rm mix}=\left(\begin{array}{cc}R&T\\ V&W\end{array}\right)\equiv
\left(\begin{array}{cc}\varkappa U&\sqrt{1-\varkappa^2}\,a\\
\sqrt{1-\varkappa^2}\,bU&\varkappa c \end{array}\right),
\label{eq_Utilde}
\end{equation}
where $b$ is also an arbitrary unitary $3\!\times\!3$ matrix ($bb^+=I$),
moreover $c=-ba$. Under these conditions the $U_{\rm mix}$ matrix will be
unitary, too ($U_{\rm mix}U_{\rm mix}^+=I$). In particular, we will use the
following $a$ and $b$ matrices:
\begin{subequations}
\begin{equation}
a=\left(\begin{array}{ccc}\,\,\,\,\,\cos\eta_2 & \sin\eta_2 & 0\\
-\sin\eta_2 & \cos\eta_2 & 0\\
\qquad 0 & 0 & e^{-i\kappa_2}\end{array}\right),\tag{3a}
\label{eq_matricesa}
\end{equation}
\begin{equation}
b=-\left(\begin{array}{ccc}\,\,\,\,\,\cos\eta_1 & \sin\eta_1 & 0\\
-\sin\eta_1 & \cos\eta_1 & 0\\
\qquad 0 & 0 & e^{-i\kappa_1}\end{array}\right),\tag{3b}
\label{eq_matricesb}
\end{equation}
\label{eq_matricesab}
\end{subequations}
where $\kappa_1$ and $\kappa_2$ are mixing phases between active and sterile
neutrinos, while $\eta_1$ and $\eta_2$ are mixing angles between them. The
remaining elements of the matrix $U_{\rm mix}$ are obtained in a standard way.

To make calculations more specific, we will use the following test values of
the new mixing parameters:
\begin{equation}
\kappa_1=\kappa_2=-\pi/2,\quad \eta_1=5^{\circ},
\quad \eta_2=\pm 15^{\circ}, \pm 30^{\circ},
\label{eq_etakappa}
\end{equation}
and restrict the values of the small parameter $\epsilon$ as
$\epsilon\lesssim 0.1$.

Let us specify the neutrino masses by the set of values
$\{m\}=\{m_i,m_{i'}\}$. For active neutrino masses, we take the estimates
presented in the works~\cite{Zysina2014,PAZH2016} for the NO case (in units of
eV), which do not contradict recent experimental data: $m_1\approx 0.0016$,
$m_2\approx 0.0088$, $m_3\approx 0.0497$. The values of mixing angles
$\theta_{ij}$ for three active neutrinos, which define the
Pontecorvo--Maki--Nakagawa--Sakata matrix, are calculated from the relations
$\sin^2\theta_{12}\approx 0.318$, $\sin^2\theta_{23}\approx 0.566$ and
$\sin^2\theta_{13}\approx 0.0222$. These relations are obtained on the basis
of processing of experimental data for the NO-case and are given in the
paper~\cite{salas}.

In order to choose the values of the masses $m_4$ and $m_5$, we use the
results of the experiments BEST, DANSS, NEUTRINO-4, MiniBooNE and MacroBooNE
\cite{ser,agu22,bar22,bar22a}. The BEST, DANSS and NEUTRINO-4 experiments are 
devoted 
to testing the existence of GA and RA associated with a deficit of electron
neutrinos and antineutrinos at short distances from the source, respectively.
It would be expected that the values of the sterile neutrino mass determined
in these two experiments would be practically the same. However, if the value
of $m_4=1.1$ in eV agrees with the data of the BEST experiment, but for the
NEUTRINO-4 experiment the analogous value is equal to $m_4=2.5$~eV. Since
further we will use one more result of the BEST experiment for the value of
$R$ (see below), we choose $m_4=1.1$~eV. For the value of $m_5$, we take
either the value $0.6$~eV close to the results of the MiniBooNE and MacroBooNE
experiments, or $0.002$~eV \cite{vstr2}. The value $m_4=1.1$~eV coincides with
the value of this parameter, which was previously used in the $(3+3)$ model
\cite{KhFom}. As for the value of $m_6$, its justified choice can be made
according to the results of future special experiments. Description of the
results of the experiments under consideration, due to the use of a specific
mixing matrix form (\ref{eq_matricesa}), do not depend in this context on a
value of $m_6$. Nonetheless we choose $m_6$ equal to $0.001$~eV following
Ref.~\cite{vstr2}.

In order to put a place for results of future experiments let us to consider
three possibilities for LSN mass values. Masses of rather heavy LSN (HLSN)
fall within the range between 0.2 and 3~eV, masses of intermediate LSN (ILSN)
between 0.003 and 0.2~eV, masses of very light SN (VLSN) between 0.0002 and
0.003~eV (see, for instance, \cite{AdeGo,vstr2,vstr1,vstr3}). Then one can
consider four  cases, namely, cases A and B, when SN mass states belong to
HLSN and VLSN ($\Lambda_i$ and $\lambda_i$ states), case C, when SN mass
states belong only to HLSN ($P_i$ states), case D, when SN mass states belong
only to VLSN ($\rho_i$ states). In this paper let us consider the A case
($\Lambda_i$ states, $i=1,2$ and $\lambda$ state, $\eta_2=\pi/6$) and the B
case ($\Lambda$ state and $\lambda_i$ states, $i=1,2$, $\eta_2=\pi/12$). We
will use in eV $\Lambda =1.1$, $\lambda_2=0.002$, $\lambda_1=0.001$,
$\Lambda_1=1.1$, $\Lambda_2=0.6$, $\lambda=0.001$. The A case uses the $m_4$
and $m_5$ mass values which are close to results based on experimental data
from Refs.~\cite{sinev,aleks}. The B case uses the $m_4$, $m_5$ and $m_6$ mass
values which are close to ones from Ref.~\cite{vstr2}.

One can generalize analytical expressions for the probabilities of transitions
and conservation of various neutrino flavors \cite{Bilenky} to the case of
decaying neutrinos. Using equations for propagation of various neutrino
flavors (see, for example,~\cite{khfo}), it is possible to obtain analytical
expressions for the transition probabilities of various flavors of stable
neutrinos/antineutrinos in a vacuum as a function of distance $L$ from the
source. If $\widetilde{U}\equiv U_{\rm mix}$ is a generalized $6\!\times\!6$
mixing matrix in the form of expression (\ref{eq_Utilde}), and if one uses
notation $\Delta_{ki}\equiv\Delta m_{ik}^2L/(4E)$, then,
following~\cite{Bilenky}, it is possible to calculate the transition
probabilities from $\nu_{\alpha}$ to $\nu_{\alpha^{\prime}}$, or from
$\overline{\nu}_{\alpha}$ to $\overline{\nu}_{\alpha^{\prime}}$ by the formula
\begin{eqnarray}
P(\nu_{\alpha}(\overline{\nu}_{\alpha})\rightarrow\nu_{\alpha^{\prime}}
(\overline{\nu}_{\alpha^{\prime}}))=\delta_{\alpha^{\prime}\alpha}
&\!\!\!\!-\,4\sum_{i>k}{\rm Re}(\widetilde{U}_{\alpha^{\prime} i}
\widetilde{U}_{\alpha i}^{\ast}\widetilde{U}_{\alpha^{\prime} k}^{\ast}
\widetilde{U}_{\alpha k})\sin^2\Delta_{ki}\,\nonumber \\
&\!\!\pm\,2\sum_{i>k}{\rm Im}(\widetilde{U}_{\alpha^{\prime} i}
\widetilde{U}_{\alpha i}^{\ast}\widetilde{U}_{\alpha^{\prime} k}^{\ast}
\widetilde{U}_{\alpha k})\sin 2\Delta_{ki}\,,
\label{eq_Bilenky}
\end{eqnarray}
\noindent where the upper sign $(+)$ corresponds to neutrino transitions
$\nu_{\alpha}\rightarrow\nu_{\alpha^{\prime}}$ while the undersign $(-)$
corresponds to antineutrino transitions
$\overline{\nu}_{\alpha}\rightarrow\overline{\nu}_{\alpha^{\prime}}$. Note
that the flavor indices $\alpha$ and $\alpha^{\prime}$ (as well as the
summation indices $i$ and $k$ over mass states) are applied to all neutrinos,
that is, to active and sterile neutrinos. Moreover, as follows from the
equation (\ref{eq_Bilenky}), the relation
$P(\nu_{\alpha}\rightarrow\nu_{\alpha})\equiv P(\overline{\nu}_{\alpha}
\rightarrow\overline{\nu}_{\alpha})$ is fulfilled exactly due to the
CPT-invariance condition~\cite{Bilenky}.

The expressions (\ref{eq_Bilenky}) given above are directly generalized to the
case of decaying neutrino mass states $k$ with including of their decay
widths $\Gamma_k$ \cite{KhFom}. To do this, it is necessary to substitute
$E_k-i\Gamma_k/2$ instead of the neutrino energy $E_k$ into the original
equations for the propagation of neutrino flavors \cite{khfo}, where
$\Gamma_k\approx m_k\Gamma_k^{(0)}/E_k$ is the decay width in the
laboratory frame, and $\Gamma_k^{(0)}$ is the same in the rest frame. Then,
the probabilities of transitions from $\nu_{\alpha}$ to
$\nu_{\alpha^{\prime}}$, or from $\overline{\nu}_{\alpha}$ to
$\overline{\nu}_{\alpha^{\prime}}$, will be calculated as follows:
\begin{eqnarray}
P(\nu_{\alpha}(\overline{\nu}_{\alpha})\rightarrow\nu_{\alpha^{\prime}}
(\overline{\nu}_{\alpha^{\prime}}))=\delta_{\alpha^{\prime}\alpha}
&\!\!\!\!\!+2\!\sum_{i>k}{\rm Re}(\widetilde{U}_{\alpha^{\prime} i}
\widetilde{U}_{\alpha i}^{\ast}\widetilde{U}_{\alpha^{\prime} k}^{\ast}
\widetilde{U}_{\alpha k})(T_{ki}\cos 2\Delta_{ki}-1)\,\nonumber \\
&\!\!\!\!\!\!\!\!\!\!\!\!\!\!\!\!\!\!\pm 2\!\sum_{i>k}{\rm Im}
(\widetilde{U}_{\alpha^{\prime} i}\widetilde{U}_{\alpha i}^{\ast}
\widetilde{U}_{\alpha^{\prime} k}^{\ast}\widetilde{U}_{\alpha k})T_{ki}
\sin 2\Delta_{ki}\,,
\label{eq23}
\end{eqnarray}
\noindent where $T_{ki}=\exp\{-\frac{L}{2}(\Gamma_k+\Gamma_i)\}\equiv
\exp\{-\frac{L}{2E}(m_k\Gamma_k^{(0)}+m_i\Gamma_i^{(0)})\}$. The upper sign
$(+)$ corresponds to neutrino transitions $\nu_\alpha\to\nu_{\alpha'}$, and
the lower sign $(-)$ corresponds to antineutrino transitions
$\bar{\nu}_\alpha\to\bar{\nu}_{\alpha'}$. For this,
$(\Delta m_{ik}^2L)/(4E\hbar c)$ is equivalent to $1.27(\Delta m_{ik}^2L)/E$,
if $\Delta m_{ik}^2$ is given in eV$^2$, $E$ is given in MeV, and $L$ is
given in meters. Respectively, $L(\Gamma_k+\Gamma_i)/2\hbar c$ is equivalent
to $0.253\cdot10^7L(\Gamma_k+\Gamma_i)$, if $L$ is given in meters, and
$\Gamma_k$ and $\Gamma_i$ are given in eV. However, $\Gamma_k$ and $\Gamma_i$
typical evaluations admit to neglect their values below.

\section{Determination of the mixing among active and sterile neutrinos with
the experiment BEST results and neutrino characteristics for some processes
with LSN contributions}
\label{ModelParameters}

\begin{figure}[htbp]
\center
\includegraphics[width=0.9\textwidth]{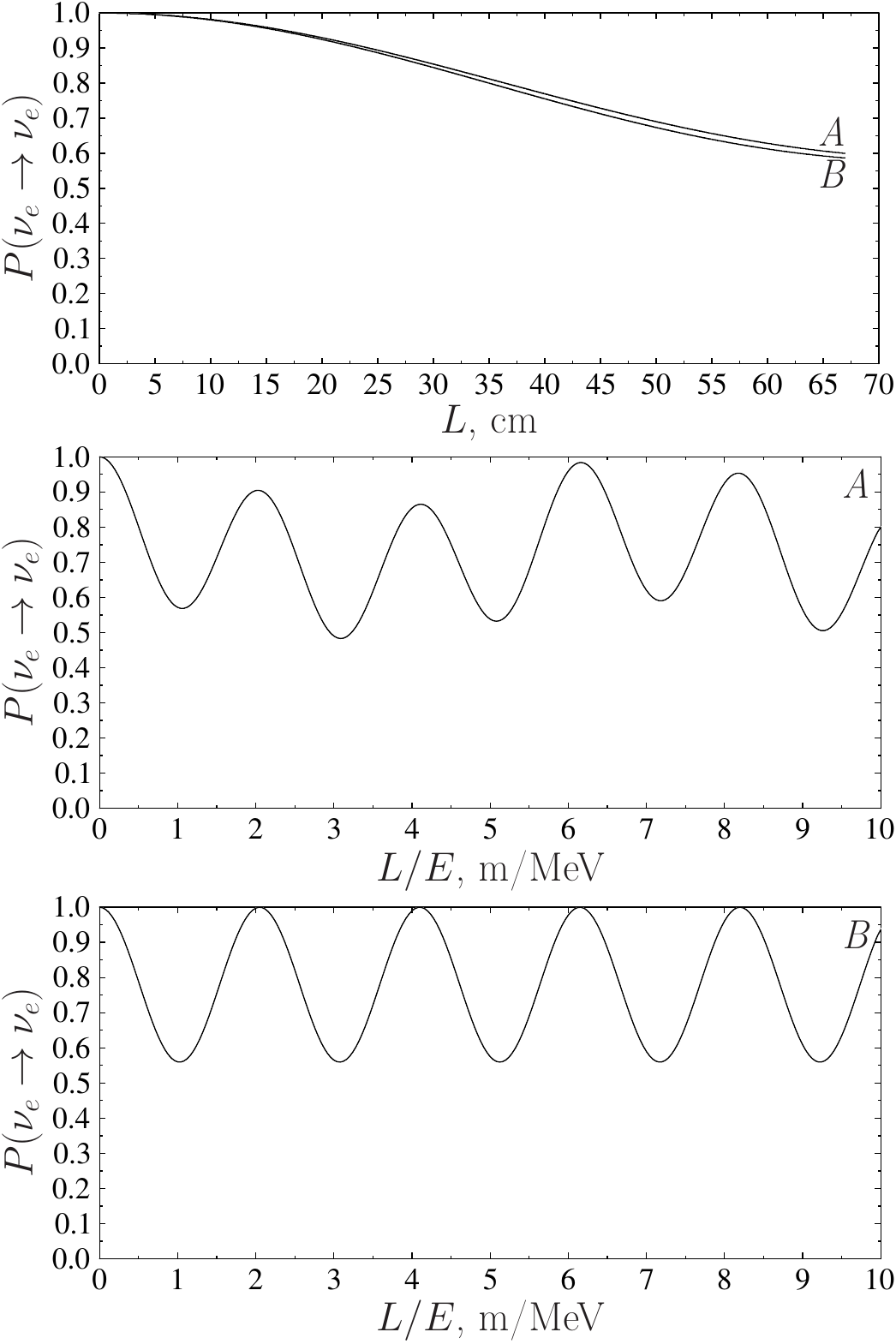}
\caption{(a) The survival probability for $\nu_e$ depending on the distance
$L$ from the source for the conditions of the BEST experiment with the values
of the parameters of the (3+3) model $m_4=1.1$~eV, $m_5=0.6$~eV,
$\eta_2=30^{\circ}$, $\epsilon=0.08$ (A), or $m_5=0.002$~eV,
$\eta_2=15^{\circ}$, $\epsilon=0.07$ (B). $L_{\rm in}=67$~cm, $R=0.8$.
(b) The survival probability for $\nu_e$ ($\bar{\nu}_e$) depending on
the ratio of the distance $L$ from the source to the neutrino energy $E$ in
the beams $\nu_{e}$ ($\bar{\nu}_e$) for sterile neutrinos in the
(3+3) model with parameter values $m_4=1.1$~eV, $m_5=0.6$~eV,
$\eta_2=30^{\circ}$, $\epsilon=0.08$ (A), or $m_5=0.002$~eV,
$\eta_2=15^{\circ}$, $\epsilon=0.07$ (B).}
\label{fig1}
\end{figure}

In the model under consideration, the value of the mixing parameter $\epsilon$
between active and sterile neutrinos is of great importance. Let's estimate
its value according to the results of experiment BEST \cite{bar22,bar22a}. In
this experiment, the detector filled with liquid gallium is divided into two
cavities. The internal cavity is a sphere with a radius of about one meter,
into which a cylinder with a radioactive source is inserted. Inside the
sphere, it can be neglected by the influence of oscillations of active
neutrinos themselves, but the effect of electron neutrino oscillations due to
mixing with LSN must be taken into account. To do this, we find the survival
probability of electron neutrinos $P_{ee}(E_i,L)$ depending on $L$ for
different values of the neutrino energy $E_i$ using the formula (\ref{eq23})
with the values of masses $m_i$ for $i=1,2,3,4,5$ given in
Section~\ref{Section_OscillationModel}. To find the value of $R$, which
characterizes the deficit of electron neutrinos for an internal target, it is
necessary to integrate over the interior of the sphere according to the
formula \cite{giut}
\begin{equation}
R\approx\frac{\int_V L^{-2}\Sigma_i P_{ee}(E_i,L)B_i\sigma_i}
{\int_V L^{-2}\Sigma_i B_i\sigma_i}.
\label{Ratio}
\end{equation}
In the formula (\ref{Ratio}), $L$ is the distance from the source to a point
inside the detector, $B_i$ are the partial ratios, $\sigma_i$ are the cross
sections of capture by $^{71}$Ga of electron neutrinos emitted from discrete
levels of $^{51}$Cr, $P_{ee}(E_i,L)$ is the survival probability of electron
neutrinos in the considered $(3+3)$ model, which is given by expression
(\ref{eq23}). Thus, an estimate of the average value of $R_{in}$ can be
obtained, which, according to \cite{bar22,bar22a}, should be equal to
experimental value of $R_{in}=0.791\pm 0.05$. At the same time, it is
necessary take into account that in $81.63\%$ of cases electron neutrinos are
emitted with an energy of 747~keV, in $8.95\%$ of cases they are emitted with
an energy of 427~keV, in $8.49\%$ of cases they are emitted with an energy of
752~keV, and in $0.93\%$ of cases they are emitted with an energy of 432~keV.
Calculations were carried out for different values of the sphere radius from
50 to 70~cm. To estimate the value of $\epsilon$, the value $L_{\rm in}=67$~cm
was chosen as the boundary value of the radius, which is the geometric radius
of the sphere. We perform calculations with various values of $\epsilon$ from
0.01 to 0.1. The most suitable for concordance with the experimental value of
$R_{in}$ are $\epsilon=0.08$ for the A case and $\epsilon=0.07$ for the B
case.
Fig.~\ref{fig1} shows the behaviour of
$P_{ee}(L)=\sum_iP_{ee}(E_i,L)B_i\sigma_i/\sum_iB_i\sigma_i$ inside the sphere
depending on the distance $L$ to the neutrino source.

\begin{figure}[htbp]
\center
\includegraphics[width=0.95\textwidth]{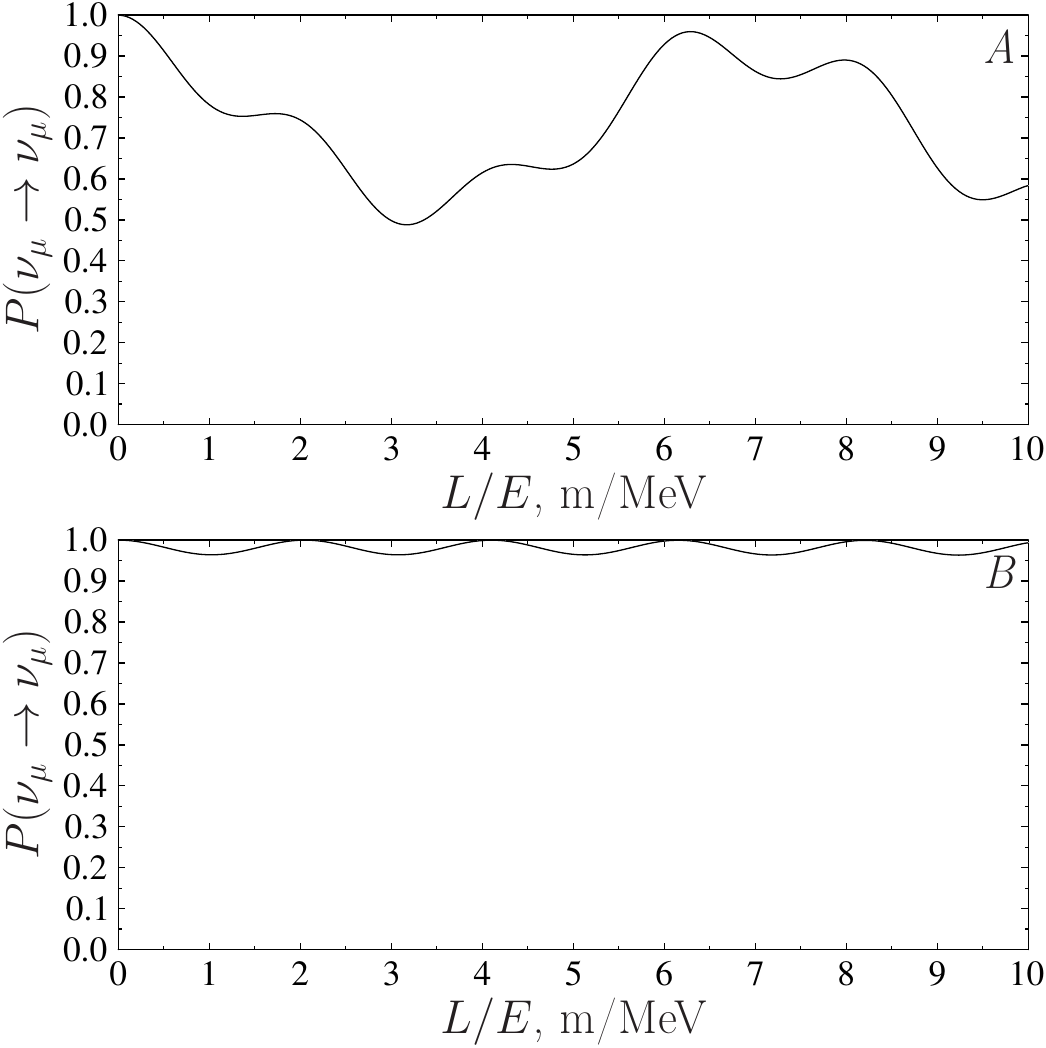}
\caption{The survival probability of $\nu_\mu$ depending on the ratio of the
distance $L$ from the source to the neutrino energy $E$ in the $\nu_{\mu}$
beams for sterile neutrinos in the (3+3) model with parameter values
$m_4=1.1$~eV, $m_5=0.6$~eV, $\eta_2=30^{\circ}$, $\epsilon=0.08$ -- top panel
(A) and $m_5=0.002$~eV, $\eta_2=15^{\circ}$, $\epsilon=0.07$ -- bottom panel
(B).}
\label{fig2}
\end{figure}
\begin{figure}[htbp]
\center
\includegraphics[width=0.95\textwidth]{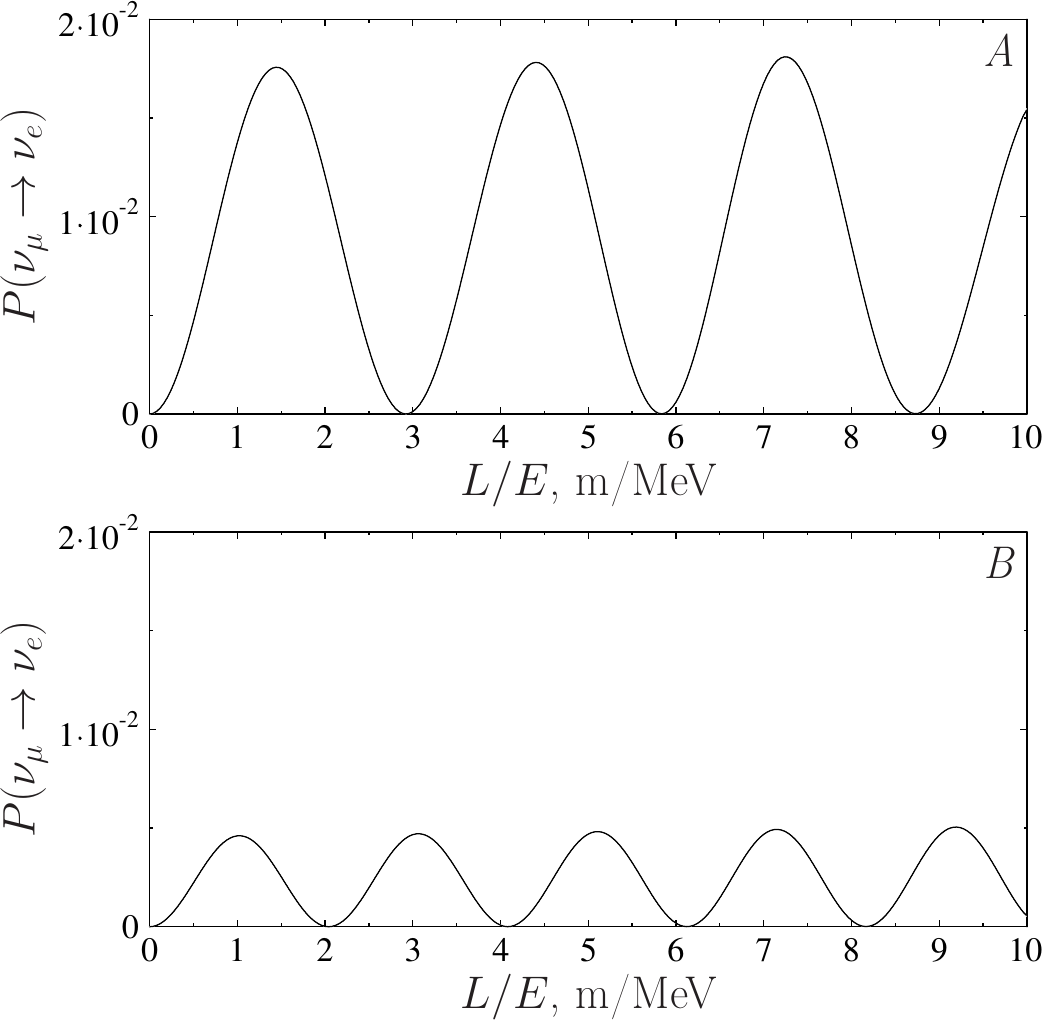}
\caption{The appearance probability of $\nu_e$ depending on the ratio of the
distance $L$ from the source to the neutrino energy $E$ in the $\nu_{\mu}$
beams for sterile neutrinos in the (3+3) model with parameter values
$m_4=1.1$~eV, $m_5=0.6$~eV, $\eta_2=30^{\circ}$, $\epsilon=0.08$ -- top panel
(A) and $m_5=0.002$~eV, $\eta_2=15^{\circ}$, $\epsilon=0.07$ -- bottom panel
(B).}
\label{fig3}
\end{figure}

After determining the value of $\epsilon$, using the values of the other
parameters given above, one can calculate and plot graphs for the survival
probability of electron neutrinos and the probability of transition of muon
neutrinos to electron neutrinos depending on the ratio $L/E$ in m/MeV, that
is, on the ratio of the distance $L$ from the source to the neutrino energy
$E$. These dependencies are important for interpreting the results of the
experiments BEST, DANSS, NEUTRINO-4, MiniBooNE and MacroBooNE 
(see Fig.~\ref{fig1}, Fig.~\ref{fig2} and Fig.~\ref{fig3}). 
The advantage of the considered model
(the A variant) is the capacity to describe processes of survival and
transition for different neutrinos types in the single model. Really in the
$(3+3)$-model in the general case three additional neutrino masses are
available instead of only one mass in the $(3+1)$-model.

One can also estimate the effective masses of the electron neutrino
$m_{\beta}$ and $m_{\beta\beta}$, which are used in calculating the
probabilities of beta decay and neutrinoless double-beta decay (the
corresponding set of Majorana phases should be used for the minimal value of
$m_{\beta\beta}$), that are for the A and B variants:
\begin{equation}
m_{\beta}=(\Sigma_i|U_{{\rm mix},ei}|^{2}m_i^2)^{1/2},\quad
m_{\beta\beta}=|\Sigma_i U^{2}_{{\rm mix},ei}m_i|.
\label{mass_bb}
\end{equation}
The obtained values in eV for $m_{\beta}\approx 0.39$ for both cases and
$m_{\beta\beta}\approx 0.1$ for case A and $m_{\beta\beta}\approx 0.13$ for
case B do not contradict with the currently available experimental results
\cite{aker,bill} for large values of the confidence probability.

\section{Discussion and conclusions}
\label{Section_Conclusion}

In this work, we have used the results of the experiments BEST, MiniBooNE and
MacroBooNE to obtain estimates of a number of parameters of the
phenomenological $(3+3)$ neutrino model with active and sterile neutrinos.
Note that these values, including LSN masses, do not have to match the
parameters of the $(3+1)$ model, and to determine them from experimental data
it needs to apply an appropriate scheme that are independent of the $(3+1)$
model. The values of parameters, for which there are currently no experimental
data, were selected as test values. If in the future the existence of light
sterile neutrinos will be reliably confirmed, this will lead to a significant
change of $\nu$SM and explanation of some phenomena in neutrino physics.
Moreover sterile neutrinos with various masses can participate in
astrophysical and cosmological processes \cite{abaza2}.

The main results of this work, which are important both for interpretation of
the data obtained in ongoing experiments and for predicting the results of
planned neutrino experiments are briefly outlined below. Based on the results
of the BEST, MiniBooNE and MacroBooNE experiments, the possible values of
masses $m_4$ and $m_5$ of sterile mass states $\nu_4$ and $\nu_5$ have been
given. According to the value of $R$, which characterizes the deficit of
electron neutrinos and measured in the BEST experiment, the main mixing
parameter $\epsilon$ between active and sterile neutrinos in the $(3+3)$
neutrino model has been estimated. The values of angles and phases of mixing
$\kappa_1$, $\kappa_2$, $\eta_1$ and $\eta_2$ are fixed as trial values.

The probability of conservation of electron and muon neutrinos and the
probability of transition of muon neutrinos to electron neutrinos depending on
the ratio of the distance $L$ from the source to the neutrino energy $E$ have
been calculated in the considered model and presented in the graphical form in
Figs 1-3. Estimates of the electron neutrino effective masses, which can be
measured in the experiments on beta decay and neutrinoless double-beta decay,
have also been also evaluated in the framework of considered model. The
important characteristic feature of the used model as compared
with the usual (3+1)-model is the possibility to describe processes with
distinct scales as $\nu_e\nu_e$, $\nu_{\mu}\nu_e$, and $\nu_{\mu}\nu_{\mu}$
processes within the unified approach.

\end{document}